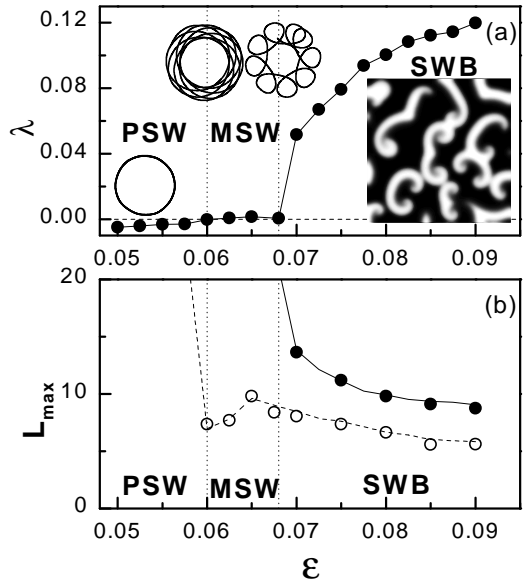

Figure 1

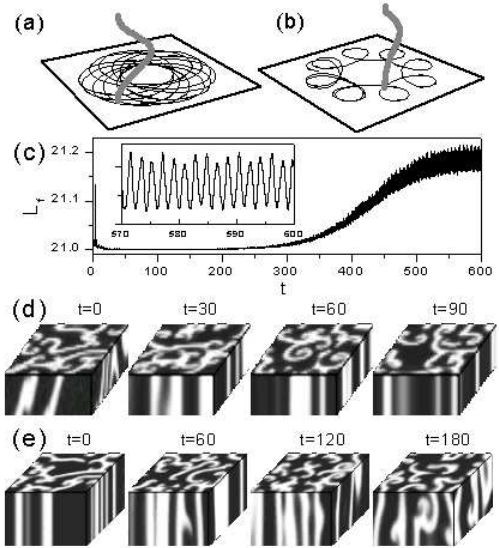

Figure 2

# Diffusion-induced Vortex Filament Instability in 3-Dimensional Excitable Media

*Zhilin Qu, Fagen Xie, and Alan Garfinkel*
Departments of Medicine (Cardiology) and Physiological Science, UCLA, Los Angeles, CA 90095

We studied the stability of linear vortex filaments in 3-dimensional (3D) excitable media, using both analytical and numerical methods. We found an intrinsic 3D instability of vortex filaments that is diffusion-induced, and is due to the slower diffusion of the inhibitor. This instability can result either in a single helical filament or in chaotic scroll breakup, depending on the specific kinetic model. When the 2-dimensional dynamics were in the chaotic regime, filament instability occurred via on-off intermittency, a failure of chaos synchronization in the third dimension.

A variety of behaviors have been observed in spiral waves in 2-dimensional (2D) excitable media [1, 2], and in scroll waves, their 3-dimensional (3D) analogs. Helical scroll waves, knotted and unknotted vortex rings, and scroll breakup behaviors, have all been observed [3-9]. However, the stability of scroll waves in 3D excitable media and its relation to the stability of 2D spiral waves, is not well understood. Recently, the stability of linear (that is, vertically straight) vortex filaments in the complex Ginzburg-Landau equation (CGLE) was investigated, showing that helical filaments and complex entangled vortices can result from an intrinsic 3D instability [4,7]. Aranson and Mitkov [8] studied the stability of vortex filaments in a reaction-diffusion system, by approximating the core dynamics using CGLE. They showed that scroll waves in 3D are more unstable than spiral waves in 2D. In this Letter, we studied the stability of linear vortex filaments in 3D reaction-diffusion systems, also using both analytical and numerical methods. We found that the intrinsic 3D instability observed in excitable media is induced by the slow diffusion of the inhibitor. We also found that in the chaotic regime, filaments lose their stability via on-off intermittency, as in chaos synchronization in other systems [10, 11].

*Stability Analysis*— Our excitable medium consists of two substances $u$ and $v$, described by a 2-variable reaction-diffusion system. The partial differential equations are [3,8]:

$$\partial u / \partial t = f(u,v) + D\nabla_{xy} u + D\nabla_z u,$$
$$\partial v / \partial t = g(u,v) + \gamma D\nabla_{xy} v + \gamma D\nabla_z v, \quad (1)$$

where $\nabla_{xy} = \partial^2/\partial x^2 + \partial^2/\partial y^2$, $\nabla_z = \partial^2/\partial z^2$, D is the diffusion constant, $\gamma \geq 0$ is the ratio of diffusion constants of $u$ and $v$, and $f$ and $g$ are given by the specific kinetic model. Assuming that a 2D spiral wave solution $\{u_0(x,y,t), v_0(x,y,t)\}$ exists, it can be periodic, quasiperiodic (meandering), or chaotic (chaotic meander and/or breakup). To analyze the stability of the corresponding scroll waves in 3D, we consider the scroll wave obtained by stacking identical 2D solutions. Then a perturbation is given to this scroll wave, i.e.,

$$u(x,y,z,t) = u_0(x,y,t) + \delta u(x,y,t)\exp(ik_z z)$$

and

$$v(x,y,z,t) = v_0(x,y,t) + \delta v(x,y,t)\exp(ik_z z).$$

Inserting this perturbation into Eq.(1), we have

$$\partial \mathbf{w}/\partial t = (\mathbf{M} - Dk_z^2 \Gamma)\mathbf{w} \quad (2)$$

where

$$\mathbf{w} = \begin{pmatrix} \delta u(x,y,t) \\ \delta v(x,y,t) \end{pmatrix}, \quad \mathbf{M} = \begin{pmatrix} f'_u + D\nabla_{xy} & f'_v \\ g'_u & g'_v + \gamma D\nabla_{xy} \end{pmatrix},$$

$\Gamma = (\Gamma_{xy})$, and $\Gamma_{xy} = \begin{pmatrix} 1 & 0 \\ 0 & \gamma \end{pmatrix}$. $\mathbf{w}$, $\mathbf{M}$, and $\Gamma$ are continuous matrices in the x-y coordinate system. Eq. (2) can be solved as:

$$\mathbf{w}(t) = \exp(\int_0^t \mathbf{M}dt - Dk_z^2 \Gamma t)\mathbf{w}(0). \quad (3)$$

Thus, the 3D stability problem of a linear vortex filament is reduced to a 2D problem, which is much easier to handle. In principle, we can numerically calculate, from Eqs. (2) or (3), the eigenvalue spectrum or Lyapunov exponent spectrum (if the 2D spiral waves are periodic), or the Lyapunov exponent spectrum (if the 2D spiral waves are aperiodic). A linear vortex filament is stable if $\mathbf{w}(t)$ in Eq. (3) is shrinking, or equivalently, if all the eigenvalues or Lyapunov exponents of this state are negative, for any nonzero $k_z$ that the system can select. This is true no matter what the stability of the 2D spiral wave is. In other words, *a filament is stable if the 2D periodic or aperiodic* *motions can be synchronized in the third dimension.* When the spiral wave is chaotic, a stable filament is the synchronized set of chaotically-moving spiral tips, i.e., spatiotemporal chaos synchronization, a subject of current interest [10,11].

When the diffusion rates of the two substances $u$ and $v$ are identical, then $\gamma=1$. $\Gamma$ then is proportional to a unitary matrix, and Eq.(3) can be rewritten as:

$$\mathbf{w}(t) = \exp(\int_0^t \mathbf{M}dt)\exp(-Dk_z^2\Gamma t)\mathbf{w}(0). \quad (4)$$

From Eq.(4), the eigenvalue spectrum (or Lyapunov exponent spectrum) of a scroll wave in 3D [$\lambda(x,y,k_z)$] is explicitly expressed in terms of the eigenvalue spectrum (or Lyapunov exponent spectrum) of spiral waves in 2D [$\lambda(x,y)$] as:

$$\lambda(x,y,k_z) = \lambda(x,y) - Dk_z^2. \quad (5)$$

Eq. (5) implies that, in the case $\gamma=1$, the stability of a linear filament can *never* be worse than the stability of the corresponding 2D spiral waves, because $\lambda(x,y,k_z) \leq \lambda(x,y)$ for any $k_z$. Thus, the filament is always stable if the corresponding 2D spiral wave is stable. If the 2D spiral wave is *un*stable, the 3D filament may become unstable in 3D when the thickness exceeds a certain critical value, leading to helical scroll waves or scroll breakup, depending on the 2D dynamics. The critical thickness for the instability in this case is:

$$L_{\max} = n\pi/k_{zc} = n\pi\sqrt{D/\lambda_{\max}} \quad (6)$$

where n=1 for no-flux boundary conditions and n=2 for periodic boundary conditions; $\lambda_{\max}$ is the maximum Lyapunov exponent or the real part of the maximum eigenvalue of the 2D spiral waves, and $k_{zc}$ is the critical wave number where the real part of $\lambda(x,y,k_z)$ changes sign from positive to negative as $k_z$ increases.

When the two components $u$ and $v$ diffuse at different rates, then $\gamma \neq 1$, Eq. (3) can not be decomposed into Eq. (4), and a simple relation like Eq. (5) is not available. In this case, the stability of the scroll wave may differ from that of the 2D spiral waves, due to the unequal diffusion rates. Instabilities due to unequal diffusion have been observed in reaction-diffusion systems [12] and in other diffusively coupled nonlinear oscillators [13]. Unlike the equal-diffusion case, it is difficult to study the stability of vortex filaments in this case analytically. Eq. (2) must be integrated to calculate the eigenvalues or Lyapunov exponents, to determine the stability of the scroll waves in 3D.

*Numerical Simulation*—Changing $\gamma$ in Eq.(1) can qualitatively change the spiral dynamics in 2D. Therefore, in order to keep the same 2D spiral wave dynamics, but study how $\gamma$ affects the stability of the vortex filaments, we removed the term $\gamma D \nabla_{xy} v$ in Eq.(1), i.e., Eq.(1) becomes:

$$\begin{aligned}\partial u/\partial t &= f(u,v) + D\nabla_{xy}u + D\nabla_z u,\\ \partial v/\partial t &= g(u,v) \quad\quad\quad + \gamma D\nabla_z v.\end{aligned} \quad (7)$$

With this change, the stability analysis above still applies, and the 2D spiral dynamics no longer change with $\gamma$. Thus Eq. (7) is valid for the purpose of linear stability analysis. When $\gamma=0$, it reduces to Eq.(1). We used the forward Euler method with $\Delta x=0.35$ and $\Delta t=0.015$ for all numerical simulations. The spatial domain in the x-y plane was fixed at $52.5^2$, with $L_z$ varying. D was set = 1. No-flux boundary conditions were used. To avoid boundary effects, the spiral core was set in the central region of the x-y plane. Since it is difficult to calculate the entire eigenvalue spectrum or Lyapunov exponent spectrum, we studied the stability by calculating the maximum Lyapunov exponent $\lambda$, using $\lambda = \lim_{t\to\infty}\frac{1}{t}\ln\frac{\|\mathbf{w}(t)\|}{\|\mathbf{w}(0)\|}$ from Eq. (2). In 3D scroll wave simulations, we used the quantity: $s(t) = \iiint_{x,y,z}[u(x,y,z,t)-\bar{u}(x,y,t)]^2 dxdydz$, to check the filament stability, where $\bar{u}(x,y,t) = \frac{1}{L_z}\int_z u(x,y,z,t)dz$. If $\lim_{t\to\infty}s(t) \to 0$, then the linear filament is asymptotically stable.

We carried out simulations using two models: model A is by Bar and Eiswirth [2]: $f(u,v)=-u(u-1)*(u-(v+b)/a)/\varepsilon$, $g(u,v)=-v$, if $u<1/3$; $g(u,v)=1-6.75u(u-1)^2-v$, if $1/3<u<1$; $g(u,v)=1-v$, if $u>1$. We fixed $a=0.84$ and $b=0.07$. In this model, increasing $\varepsilon$ results in a series of bifurcations, from a periodic spiral wave to a meandering spiral wave and finally to spiral wave breakup [2] [Fig.1(a)]. Figure 1(a) also shows $\lambda$ versus $\varepsilon$ for the corresponding spiral wave dynamics. Note that the behavior becomes chaotic ($\lambda > 0$) beyond a critical value of $\varepsilon$. Figure 1(b) shows the maximum thickness $L_{\max}$ below which linear vortex filaments are stable, from 2D Lyapunov exponent analysis and from direct 3D scroll wave simulations. In 3D simulations, we stacked up 2D spiral waves in the z-direction and gave a small random perturbation to this initial condition. We then monitored $s(t)$ to observe whether the perturbation decayed to zero or expanded. Note that the 3D simulation results agree very well with the predictions from 2D Lyapunov analysis. $L_{\max}$ is larger for $\gamma=1$ than for $\gamma=0$, indicating that filaments are more stable for $\gamma=1$ than for $\gamma=0$.

Note in Fig.1 that when γ=1 in this model, unstable linear filaments can be seen only in the 2D spiral wave breakup regime, because λ can be positive only when the 2D dynamics is chaotic. In the chaotic regime, when $L_z>L_{max}$, synchronization of spatiotemporal chaos is lost, and 2D chaos is converted into true 3D chaos. In the meandering regime and in the periodic regime, filaments are stable for any $L_z$ because λ in Eq.(6) is never positive. When γ=0, the linear filament is stable in the periodic spiral wave regime except for a very narrow range next to the meander regime (Figure 1(b)). In the meander regime, the linear filament loses its stability when $L_z>L_{max}$, leading to a helical filament *which still meandered as in 2D* [Figs.2 (a) and (b)]. The bifurcation that underlies this loss of filament stability appears to be a form of Hopf bifurcation, since a new frequency is introduced. Fig.2(c) shows the filament length $L_f$ versus time for the filament in Fig.2(a). $L_f$ increased slowly at first and saturated at t close to 600, but it oscillates throughout with a frequency $f \approx 0.5$. This frequency is different from the rotating frequency ($f \approx 0.225$) or the meandering frequency ($f \approx 0.04$) of the 2D spiral wave.

In the spiral wave breakup regime, though the number and motions of filaments change with time chaotically, linear filaments are stable when $L_z<L_{max}$ [Fig.2(d)]. But when $L_z>L_{max}$, linear filaments lose their stability and filament bending, twisting, and breakup occur. If $L_z$ is larger than another critical thickness, scroll waves can rotate through the z-axis, leading to fully-developed 3D 'turbulence' [Fig.2(e)]. In other words, when $L_z<L_{max}$, spatiotemporal chaos synchronization occurred, but when $L_z>L_{max}$, spatiotemporal chaos synchronization is lost, and 2D chaos becomes true 3D chaos. It is interesting to note that in the spiral wave breakup regime, filaments lose their stability via 'on-off intermittency' (Fig.3), just as in spatiotemporal chaos synchronization in other systems [10,11]. The intermittent feature of s(t) shown in Fig 3(b) is due to occasional filament bending and breaking.

In Model A, this 3D instability does not extensively change the basic dynamics. A very different feature was observed in a model developed by Aliev and Panfilov [9, 14] (We refer to it as Model B): $f(u,v)=-\alpha u(u-\beta)(u-1)-uv$; $g(u,v)=[\sigma+\mu_1 v/(u+\mu_2)][-v-\alpha u(u-\beta-1)]$, where β=0.15, σ=0.002, $\mu_1$=0.2, $\mu_2$=0.3, and α=40. Gray and Jalife [9] showed, for γ=0 in this model, that a spiral wave in 2D was stable but the corresponding scroll wave in 3D was unstable, leading to chaotic scroll breakup when the thickness exceeds a certain value. In Fig.4(a), we show this scroll breakup for $L_z$=35 and γ=0. A single vertical scroll wave with a small perturbation became multiple scroll waves turning in all three axes, and displaying chaotic behavior. However, when we set γ=1, this instability disappeared [Fig.4(b)].

Our study has thus shown that the 3D vortex filament instability is diffusion-induced. This instability has different consequences in different models. In Fig.5 we show λ versus $k_z$ for the two models simulated above. For γ=0, λ increases with increasing $k_z$, from a slightly negative number [15] to positive, and then changes to negative at a critical wave number $k_{zc}$, when the 2D spiral wave is non-chaotic. But in the chaotic regime, with increasing $k_z$, λ decreases monotonically, from positive to negative at $k_{zc}$ [inset of Fig.5(a)]. In all the cases we studied, we found that the dispersion curves were lowered or depressed as γ increased. This indicates that the instability is induced by the slow diffusion of the inhibitor. Although in the non-chaotic regime, both models we studied have similar dispersion relations, i.e, similar linear stability properties, their nonlinear dynamics are different. In Model A, the diffusion-induced instability causes a straight filament of a periodic or meandering scroll wave to become a helical scroll wave, while preserving the qualitative 2D meandering behaviors [Figs.2(a) & (b)]. But in Model B, this instability caused scroll wave breakup, producing chaos [Fig.4(a)].

In this study we found that the 3D vortex instability occurs when the inhibitor diffuses slower than the activator, which is the opposite of the instability described by Turing [12], for which the inhibitor must diffuse faster than the activator. (In this Letter, we have not studied the effects of a fast inhibitor on the stability of a scroll wave; it may induce additional instabilities, as shown in [16].) The present instability may be similar to a *phase instability* in CGLE [17], occurring when the complex part of the diffusion constant is non-zero, which is the analog of unequal diffusion in excitable media [5]. Recent studies on vortex stability in the CGLE by Aranson et al [4] and Nam et al [7] found similar instabilities, leading to helical filaments and scroll breakup, when the complex part of the diffusion is nonzero. Aranson and Mitkov [8] studied the filament stability in excitable media, using the 1D CGLE approximation, which is valid only at the Hopf bifurcation point. Our study is applicable to all the dynamical regimes. An important aspect of our study compared to the CGLE analysis is that we showed that it is the *slow* diffusion of the inhibitor causing the instability. Thus, this instability may well be relevant to cardiac conduction (for which the 'inhibitor' does not diffuse at all, i.e., γ=0). In addition, we addressed the issue of critical thickness, which is practically important for cardiac conduction. Our analysis shows that the 3D instability occurs beyond a critical thickness which is determined by the intrinsic dynamics of the specific system, not the thickness required for reentry in the third dimension suggested by Winfree [18].

This research was supported by NIH SCOR in Sudden Cardiac Death P50 HL52319 and by Fellowships from the American Heart Association, Western States Affiliate to Z. Q. and F.X.

**FIGURE CAPTIONS:**

Fig.1. (a) $\lambda$ versus $\varepsilon$ for Model A. The three regions separated by dotted lines are the regions of periodic spiral waves (PSW), meandering spiral waves (MSW), and spiral wave breakup (SWB). Insets are tip trajectories of a PSW($\varepsilon$=0.055) and two MSWs ($\varepsilon$=0.06 & 0.068), and a snapshot of SWB ($\varepsilon$=0.075). (b) $L_{max}$ versus $\varepsilon$ for Model A. Lines are predictions from 2D Lyapunov analysis. Symbols are direct 3D scroll simulation results. $\gamma$=1 ( —, • ); $\gamma$=0 (----, O). The solid line is from Eq.(6) by using $\lambda$ in Fig.1(a). The dashed line was obtained by $L_{max}=\pi/k_{zc}$, where $k_{zc}$ is the critical wave number at which $\lambda$ changes from positive to negative, and $\lambda$ was obtained by integrating Eq.(2) for $\gamma$=0.

Fig.2. (a) and (b) Tip trajectories on the bottom surfaces and filaments for $L_z$=21. (a) $\varepsilon$=0.06; (b) $\varepsilon$=0.068. (c) Filament length $L_f$ versus time for the filament shown in (a); The inset shows $L_f$ for t=570 to 600. (d) and (e) Snapshots of surfaces of 3D simulations for $\varepsilon$=0.075: (d) $L_z$=6.3; (e) $L_z$=21.

Fig.3 s(t) for $\varepsilon$=0.075. (a) $L_z$=6.3; (b) $L_z$=7.35; (c) $L_z$=21.

Fig.4 Snapshots of surfaces of 3D simulations (left), and u versus t (right) of Model B for $L_z$=35. (a) $\gamma$=0; (b) $\gamma$=1. (a) and (b) started from the same initial condition.

Fig.5. $\lambda$ versus $k_z$. Full circles are $\lambda$ calculated using Eq.(2), and the thick lines for $\gamma$=1 are from Eq.(6). (a) Model A for $\varepsilon$=0.075. The insert in (a) is for $\varepsilon$=0.068. (b) Model B.

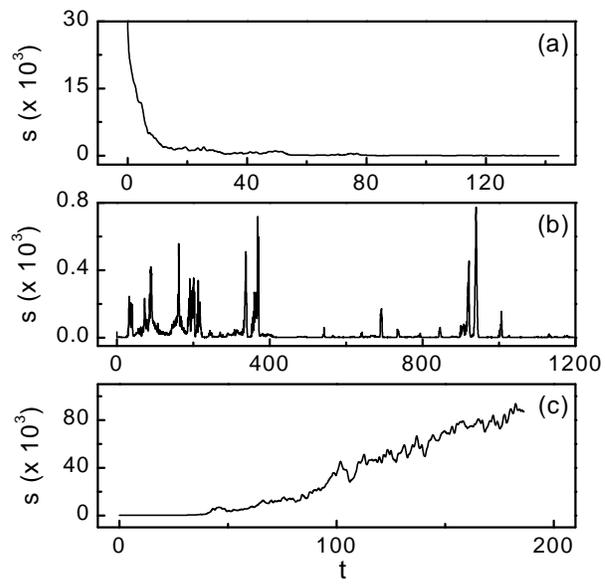

Figure 3

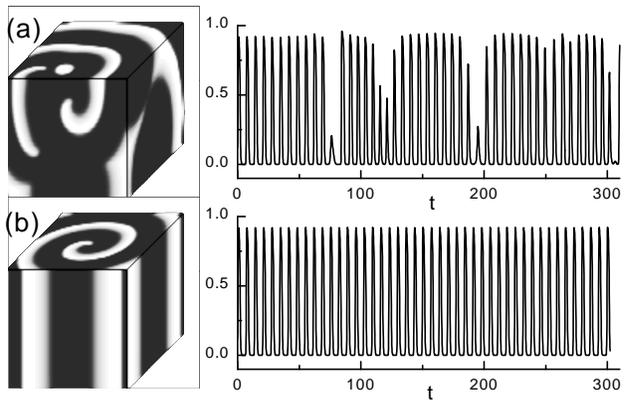

Figure 4

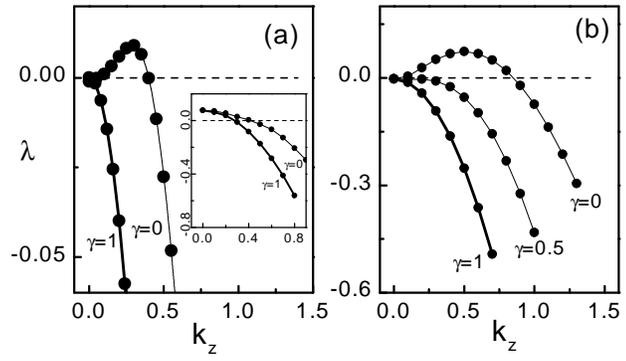

Figure 5